\documentclass[twoside,draft]{article}

\usepackage{pgfplots}
\usepackage{amsfonts}
\usepackage{stmaryrd}
\usepackage{amssymb}
\usepackage{euscript}
\usepackage{amsthm}
\usepackage{amsmath}
\usepackage{booktabs}
\usepackage{multirow}
\usepackage{array}

\usepackage{latexsym}
\usepackage{mathrsfs}
\usepackage{graphicx}
\usepackage{color}
\usepackage{dsfont}
\usepackage{bm}

\usepackage{enumitem}%

\usepackage[hidelinks]{hyperref}

\numberwithin{equation}{section}

\newtheorem{theorem}{Theorem}[section]
\newtheorem{lem}{Lemma}[section]
\newtheorem{pro}{Proposition}[section]
\newtheorem{cor}{Corollary}[section]
\newtheorem{rem}{Remark}[section]
\newtheorem{rems}{Remarks}[section]
\newtheorem{ex}{Example}[section]
\newtheorem{defi}{Definition}[section]
\newtheorem{hyp}{Assumption}[section]

\newcommand{\bt}{\begin{theorem}}
\newcommand{\et}{\end{theorem}}
\newcommand{\bl}{\begin{lem}}
\newcommand{\el}{\end{lem}}
\newcommand{\bp}{\begin{pro}}
\newcommand{\ep}{\end{pro}}
\newcommand{\bcor}{\begin{cor}}
\newcommand{\ecor}{\end{cor}}

\newcommand{\bd}{\begin{defi} \rm }
\newcommand{\ed}{\end{defi}}
\newcommand{\brem }{\begin{rem} \rm }
\newcommand{\erem }{\end{rem}}
\newcommand{\brems }{\begin{rems} \rm }
\newcommand{\erems }{\end{rems}}
\newcommand{\bhyp }{\begin{hyp} \rm }
\newcommand{\ehyp }{\end{hyp}}
\newcommand{\bex}{\begin{ex} \rm }
\newcommand{\eex}{\end{ex}}

\usepackage{comment}

\newcommand{\esssup}{\operatornamewithlimits{ess\,sup}}

\newcommand{\cS}{\mathcal{S}}

\newcommand{\cF}{{\mathcal F}}

\newcommand{\cM}{{\mathcal M}}

\newcommand{\FF}{{\mathbb F}}

\newcommand{\EE}{{\mathbb E}}

\begin{document}

\title{A Monotone Limit Approach to Entropy‑Regularized American Options \vskip35pt}


\author{Daniel Chee$\,^{a}$, Noufel Frikha$\,^{b}$ and Libo Li$\,^{a}$ \\ \\ \\ \\
\\ $^{a\,}$School of Mathematics and Statistics, University of New South Wales \\ Sydney, NSW 2052, Australia \\ \\
$^{b\,}$Universit\'e Paris 1 Panth\' eon-Sorbonne, Centre d’Economie de la Sorbonne, \\106 Boulevard de l’H\^opital, 75642 Paris Cedex 13, France\\ }

\maketitle
\vskip20pt

\begin{abstract}
Recent advances in continuous-time optimal stopping have been driven by entropy-regularized formulations of randomized stopping problems, with most existing approaches relying on partial differential equation methods. In this paper, we propose a fully probabilistic framework based on the Doob-Meyer-Mertens decomposition of the Snell envelope and its representation through reflected backward stochastic differential equations. We introduce an entropy-regularized penalization scheme yielding a monotone approximation of the value function and establish explicit convergence rates under suitable regularity assumptions. In addition, we develop a policy improvement algorithm based on linear backward stochastic differential equations and illustrate its performance through a simple numerical experiment for an American-style max call option.
\end{abstract}

\newpage

\tableofcontents
\newpage
\section{Introduction}
The numerical resolution of optimal stopping problems using Monte Carlo and machine learning techniques has been the subject of extensive research; see, for instance,
\cite{BCJ2019, BCJW2021, BCJ2020, BG2004, R2002, RST2022, STG2023, DD2024, DSXZ2024, D2023, DFX2024}.
More recently, advances in reinforcement learning (RL) have stimulated renewed interest in continuous-time formulations of optimal stopping problems arising in mathematical finance.
A particularly remarkable and recent development in this direction is the entropy-regularized randomized stopping framework introduced independently by Dong~\cite{D2023}, Dianetti \emph{et al.}~\cite{DFX2024}, and Dai \emph{et al.}~\cite{DSXZ2024}.

The entropy-regularization paradigm recasts the classical optimal stopping problem as an exploratory control problem, thereby enabling the design of RL algorithms through a partial differential equation (PDE) approach, typically via Hamilton--Jacobi--Bellman (HJB) equations.
A central feature of this framework is a policy-based interpretation, in which the control represents the instantaneous stopping intensity, conditional on survival.
This viewpoint has proved remarkably flexible and has since served as the foundation for a growing body of work extending the original formulation of~\cite{D2023}.

Among recent contributions, Dai \emph{et al.}~\cite{DDJZ2023} propose a recursive entropy-regularization scheme with biased Gaussian exploration to learn Merton-type strategies in incomplete markets.
Dai and Dong~\cite{DD2024} investigate investment problems with transaction costs through randomized Dynkin games, while Dong and Zheng~\cite{DZ2025} study mean--variance stopping problems using an extended HJB system combined with vanishing regularization.
Further extensions include the continuous-time RL framework with regime switching developed by Huang \emph{et al.}~\cite{HLYZ2025}, as well as actuarial applications to insurance surrender decisions by Jia \emph{et al.}~\cite{JWW2024}.
Collectively, these works underscore the breadth and effectiveness of entropy-regularized RL methods in continuous-time stochastic control.

The present work is motivated by this expanding literature on RL for continuous-time control
\cite{TZZ2021, WZZ2020}
and is particularly aligned with the objectives of~\cite{D2023}.
As in~\cite{D2023, DFX2024, DSXZ2024}, our starting point is the randomized stopping representation of the value process $V$ associated with an optimal stopping problem with payoff $P$.
Specifically, following Gy\"ongy and \v{S}i\v{s}ka~\cite[Theorem~2.1]{GS2008}, one has
\begin{align}
V_t
= \esssup_{\tau \in \mathcal{T}_{t,T}} \mathbb{E}[P_\tau \mid \mathcal{F}_t]
= \esssup_{\gamma \in \Lambda}
\mathbb{E}\Big[
P_T e^{-\int_t^T \gamma_u du}
+ \int_t^T P_s \gamma_s e^{-\int_t^s \gamma_u du} ds
\,\Big|\, \mathcal{F}_t
\Big],
\label{rstopping}
\end{align}
where $\mathcal{T}_{t,T}$ denotes the set of stopping times taking values in $[t,T]$ and
$\Lambda = \cup_{n=1}^\infty \Lambda_n$, with
$$
\Lambda_n := \{\gamma:\ \gamma \text{ is non-negative, adapted, and } 0 \leq \gamma \leq n\}.
$$

In contrast to the PDE-based approaches adopted in
\cite{D2023, DFX2024, DSXZ2024},
we pursue here a purely probabilistic perspective, in the spirit of our earlier work on Bermudan options in Chee \emph{et al.}~\cite{CFL2025}.
Our analysis relies on the Doob-Meyer-Mertens decomposition of the Snell envelope $V$ and its characterization through reflected backward stochastic differential equations (RBSDEs or reflected BSDEs).
Formally applying It\^o's formula to the right-hand side of~\eqref{rstopping} suggests the backward representation
\begin{gather}
V_t
= P_T - (M_T - M_t)
+ \esssup_{\gamma \in \Lambda} \int_t^T (P_s - V_s)\gamma_s \, ds,
\label{rstoppingbsde}
\end{gather}
where $M$ is a martingale.

The backward equation~\eqref{rstoppingbsde} is ill-posed, since the optimal control, if it exists, formally satisfies
$\gamma_s^\ast = \infty \cdot \mathbf{1}_{\{s \geq \tau_\ast\}}$,
where $\tau_\ast$ denotes the optimal stopping time in~\eqref{rstopping}.
To overcome this difficulty, we introduce an entropy-based penalization and consider the family of BSDEs
\begin{gather}
v^\lambda_t
= P_T - (m^\lambda_T - m^\lambda_t)
+ \esssup_{\gamma \in \Lambda}
\int_t^T \Big( (P_s - v^\lambda_s)\gamma_s
- \lambda \, (\text{penalty term}) \Big) ds,
\label{penalty}
\end{gather}
where $\lambda \geq 0$ is a temperature parameter.

Unlike~\cite{D2023}, which employs the penalty function $x\ln x - x$, we adopt the modified entropy penalty $x\ln x - x + 1$.
This choice yields a monotone sequence of BSDE solutions as $\lambda \downarrow 0$, and we therefore refer to the resulting procedure as an \emph{entropy-regularized penalization scheme}.
In contrast to the classical penalization approach for RBSDEs, which relies on truncating the control $\gamma$ (see \cite{EKPPQ1997}), our method provides a smooth and analytically tractable regularization of the reflected constraint. 

For completeness, we note that a related but distinct approach is developed in our companion working paper~\cite{CFL2026}, where entropy regularization is introduced through the classical truncation-based penalization scheme for reflected BSDEs, leading to a reflected equation with a singular driver.
The focus of~\cite{CFL2026} is on well-posedness and structural properties of the resulting RBSDE, whereas the present paper emphasizes convergence rates, policy improvement algorithms (PIA), and numerical implementation within the entropy-regularized framework.

The rest of the paper is organised as follows.
In Section~\ref{sec:preliminaries}, we introduce the probabilistic framework and recall several results from stochastic analysis that will be used throughout the paper.

Section~\ref{sec:entropy:regularized:scheme} is devoted to the entropy-regularized formulation of the optimal stopping problem.
We introduce the penalized BSDE associated with the randomized stopping representation and establish its well-posedness. Exploiting monotonicity arguments developed in El Karoui \emph{et al.}~\cite{EKPPQ1997} and Peng~\cite{P1999}, we then show that, as the temperature parameter $\lambda$ tends to zero, the entropy-regularized value process converges monotonically to the value of the original optimal stopping problem.
Under additional regularity assumptions on the payoff process, we further derive quantitative convergence rates for both lower and upper bounds. We conclude this section by studying the PIA associated with the entropy-regularized problem.
The algorithm is based on an iterative sequence of linear BSDEs and admits an explicit policy update at each step.
We prove that the resulting sequence of value functions converges increasingly to the entropy-regularized value function at a factorial rate.

Section~\ref{sec:numerics} is devoted to a simple numerical experiment. We present an efficient implementation of the PIA in the case of max-call option in the Black-Scholes setting.
Our numerical results confirm the theoretical convergence properties and demonstrate that, for small values of $\lambda$, the entropy-regularized approach provides accurate approximations of classical optimal stopping prices.

Finally, we refer interested readers to our working paper \cite{CFL2026}, which introduces entropy-regularization via the standard penalization scheme for RBSDEs, investigates the well-posedness properties of the resulting RBSDE with singular driver, and provides a probabilistic interpretation of the approach.

\section{Preliminaries}\label{sec:preliminaries}
We work on a standard filtered probability space $(\Omega, \mathcal{F}, \mathbb{P}, \mathbb{F})$, where the filtration $\mathbb{F} = (\mathcal{F}_t)_{t \geq 0}$ is assumed to satisfy the usual conditions. Any additional assumptions imposed on the filtration $\mathbb{F}$ will be stated explicitly when required. We denote by $\mathcal{O}(\mathbb{F})$ the space of $\mathbb{F}$-optional processes, by $\mathcal{P}(\mathbb{F})$ the space of $\mathbb{F}$-predictable processes, by $\mathcal{M}$ the space of $\mathbb{F}$-martingales, and by $\mathcal{A}^+$ the space of $\mathbb{F}$-predictable, non-decreasing processes. We further adopt the standard notation $x \vee y := \max(x,y)$, $x^{+} := \max(x,0)$, and $x^{-} := \max(-x,0)$. Throughout the paper, $C$ and $K$ denote generic positive constants whose values may change from line to line.

We shall work extensively with the following function spaces. The Banach space of square-integrable optional processes is defined as
$$
\mathcal{S}^2 := \big\{ X \in \mathcal{O}(\mathbb{F}) : \mathbb{E}\big[\sup_{0 \leq t \leq T} |X_t|^2\big] < \infty \big\}.
$$
The space of square-integrable $\mathbb{F}$-martingales is given by
$$
\mathcal{H}^2 := \big\{ M \in \mathcal{M} : \mathbb{E}\big[[M]_T\big] < \infty \big\},
$$
where $[M]$ denotes the quadratic variation of $M$. The space of square-integrable, $\mathbb{F}$-predictable, increasing processes is
$$
\mathcal{K}^2 := \big\{ A \in \mathcal{A}^+ : \mathbb{E}\big[A_T^2\big] < \infty \big\}.
$$

We recall that if the payoff process $P$ is càdlàg and satisfies appropriate integrability conditions, then the value process $V = (V_t)_{0 \leq t \leq T}$ associated with the corresponding optimal stopping problem belongs to class (D), that is, the family $\{V_\tau : \tau \text{ stopping time}\}$ is uniformly integrable. In this setting, $V$ admits a Doob--Meyer decomposition and equivalently satisfies the RBSDE
\begin{align}
V_t &= P_T - (M_T - M_t) + (A_T - A_t), \qquad t \in [0,T], \label{VRBSDE} \\
V_t &\geq P_t, \quad \text{and} \quad \int_0^T (V_{s-} - P_{s-})\, dA_s = 0, \nonumber
\end{align}

\noindent where $M$ is a uniformly integrable $\mathbb{F}$-martingale and $A$ is a $\mathbb{F}$-predictable, increasing process. In particular, if $P \in \mathcal{S}^2$, then $(V,M,A) \in \mathcal{S}^2 \times \mathcal{H}^2 \times \mathcal{K}^2$; see, for instance, Steps~1--4 in the proof of Lemma~3.3 in Grigorova \emph{et al.}~\cite{GIOOQ2017}.

For further background on optimal stopping problems and on the theory of BSDEs and RBSDEs under general filtrations, we refer the reader to Maingueneau~\cite{M1978}, El~Karoui \emph{et al.}~\cite{EKPPQ1997}, Øksendal and Zhang~\cite{OZ}, Grigorova \emph{et al.}~\cite{GIOOQ2017}, and Hamadène and Ouknine~\cite{HO2015}.

\vskip20pt

\section{Entropy-Regularized Penalization Scheme}\label{sec:entropy:regularized:scheme}

In this section, we exploit the connection between the randomized stopping representation \eqref{rstopping} and the associated BSDE formulation \eqref{rstoppingbsde} to introduce an entropy-regularized penalization scheme. We analyze its convergence to the price of the corresponding American option as the temperature parameter $\lambda$ tends to zero. We also introduce the associated policy improvement algorithm (PIA) and investigate its convergence properties.

\subsection{Definition of the entropy-regularized penalization scheme}

We begin by stating the standing assumption on the payoff process.

\begin{hyp}\label{assumption1}
The payoff $P$ is càdlàg and uniformly bounded, with bound denoted by $|P|_\infty$.
\end{hyp}

As previously discussed, our main idea is to penalize the control $\gamma$ directly at the level of the BSDE representation \eqref{rstoppingbsde}, rather than at the level of the randomized stopping problem itself. More precisely, we consider the BSDE
\begin{gather*}
v^\lambda_t
= P_T - (m^\lambda_T - m^\lambda_t)
+ \esssup_{\gamma \in \Lambda}
\int_t^T \Big( (P_s - v^\lambda_s)\gamma_s
- \lambda \big(\gamma_s \ln \gamma_s - \gamma_s + 1\big) \Big)\, ds.
\end{gather*}

The first-order optimality condition associated with this control problem is given by
\begin{equation}
(P_s - v^\lambda_s) - \lambda \ln(\gamma_s) = 0,
\label{firstorder}
\end{equation}
which yields the explicit expression for the optimal control (or policy)
\begin{equation}
\gamma_s = \exp\big((P_s - v^\lambda_s)/\lambda\big).
\label{dongpolicy}
\end{equation}
In particular, this optimal policy does not explode at infinity. 

Substituting \eqref{dongpolicy} back into the BSDE, we obtain the following nonlinear equation:
\begin{equation}
v^\lambda_t
= P_T - (m^\lambda_T - m^\lambda_t)
+ \int_t^T \lambda \big( e^{(P_s - v^\lambda_s)/\lambda} - 1 \big)\, ds.
\label{dongbsdemod}
\end{equation}

\newpage
Throughout the remainder of this section, we refer to the family
$$
\big\{ v^\lambda := (v^\lambda_t)_{t \in [0,T]} \;:\; \lambda \in (0,1] \big\}
$$

\noindent as the \emph{entropy-regularized penalization scheme}. We first establish the well-posedness of the BSDE \eqref{dongbsdemod}. Its convergence to the value process $V$ of the American option with payoff $P$ as $\lambda \downarrow 0$ is addressed in Section~\ref{subsec:conv:american:option}.

\begin{rem}
There are several possible ways to penalize the original control problem in \eqref{rstoppingbsde}. For instance, following \cite{D2023}, one may consider the entropy term $x \ln x - x$, which leads to the BSDE
\begin{gather}
V^\lambda_t
= P_T - (M^\lambda_T - M^\lambda_t)
+ \esssup_{\gamma \in \Lambda}
\int_t^T \Big( (P_s - V^\lambda_s)\gamma_s
- \lambda (\gamma_s \ln \gamma_s - \gamma_s) \Big)\, ds.
\label{BSDEgamma}
\end{gather}
Assuming the existence of a solution $(V^\lambda, M^\lambda) \in \mathcal{S}^2 \times \mathcal{H}^2$, the associated optimal policy is again given by \eqref{dongpolicy}, and the BSDE reduces to
\begin{equation}
V^\lambda_t
= P_T - (M^\lambda_T - M^\lambda_t)
+ \int_t^T \lambda e^{(P_s - V^\lambda_s)/\lambda} \, ds.
\label{dongbsde}
\end{equation}
\end{rem}

The above discussion illustrates that the choice of penalization is not unique and may lead to different optimal policies and families of BSDEs. However, as we shall show below, a key advantage of our formulation compared to that of \cite{D2023} is that the family of solutions $(v^\lambda)_{\lambda \in (0,1]}$ defined by \eqref{dongbsdemod} is monotone increasing as $\lambda$ decreases.

We observe that, in the present setting, the driver in \eqref{dongbsdemod} is given by
$$
x \longmapsto \lambda \big( e^{(P_s - x)/\lambda} - 1 \big),
$$

\noindent which is monotone decreasing in $x$. As a consequence, uniqueness follows from arguments similar to those developed in Lepeltier \emph{et al.}~\cite{LMX2005}. Therefore, it suffices to establish the existence of a solution, noting that the driver is only locally Lipschitz continuous.

\bp
For any $\lambda \in (0,1]$, there exists a unique solution $(v^\lambda, m^\lambda) \in \cS^2 \times \cM^2$ to \eqref{dongbsdemod}.
\ep

\begin{proof}
To prove existence, we proceed by truncation. Fix $n \geq T$ and consider the BSDE with Lipschitz continuous driver
\begin{gather*}
v^{\lambda,n}_t
= P_T - \int_t^T dm^{\lambda,n}_s
+ \int_t^T \lambda \Big( e^{(P_s - v^{\lambda,n}_s \vee (-n))/\lambda} - 1 \Big) \, ds.
\end{gather*}
By standard results for BSDEs with Lipschitz drivers (see, for example, Theorem~3.1 in \O ksendal and Zhang~\cite{OZ}), there exists a solution
$(v^{\lambda,n}, m^{\lambda,n}) \in \cS^2 \times \cM^2$.
Moreover, taking conditional expectations yields
\begin{gather*}
v^{\lambda,n}_t + \lambda (T - t)
= \EE\Big[ P_T + \int_t^T \lambda
e^{(P_s - v^{\lambda,n}_s \vee (-n))/\lambda} \, ds
\;\Big|\; \cF_t \Big]
\geq 0.
\end{gather*}
It follows that $v^{\lambda,n}_t \geq -\lambda T \geq -T \geq -n$, and therefore the truncation is inactive. Consequently,
\begin{gather*}
v^{\lambda,n}_t
= \EE\Big[ P_T + \int_t^T \lambda \big( e^{(P_s - v^{\lambda,n}_s)/\lambda} - 1 \big)\, ds
\;\Big|\; \cF_t \Big],
\end{gather*}
which shows that $(v^{\lambda,n}, m^{\lambda,n})$ actually solves \eqref{dongbsdemod}. The result then follows from uniqueness.
\end{proof}

\newpage
\subsection{Convergence to the American Option}\label{subsec:conv:american:option}

In this subsection, we show that as $\lambda \downarrow 0$, the entropy-regularized penalization scheme $(v^\lambda)_{\lambda \in (0,1]}$ converges to the value process $V$ of the American option defined in \eqref{rstopping}. We also investigate the associated rate of convergence.

To this end, we introduce the process
\begin{gather}
\gamma^\lambda_t := \frac{\lambda}{P_t - v^\lambda_t}\big(e^{(P_t - v^\lambda_t)/\lambda} - 1\big), \quad t \in [0,T]. \label{gammaintensity}
\end{gather}
With this notation, the BSDE \eqref{dongbsdemod} can be rewritten in the form
\begin{align*}
v_t^\lambda
= P_T - (m^\lambda_T - m^\lambda_t)
+ \int_t^T (P_s - v^\lambda_s)\,\gamma^\lambda_s \, ds.
\end{align*}

Observe that the function $x \mapsto x^{-1}(e^{x} - 1)$ is positive and strictly increasing on $\mathbb{R}$. As a consequence, for any fixed $\lambda \in (0,1]$, the process $\gamma^\lambda$ defined in \eqref{gammaintensity} admits a natural interpretation as a stopping intensity.

We next show that the stopping intensity process $\gamma^\lambda$ is uniformly bounded. Taking $\cF_t$-conditional expectations in \eqref{dongbsdemod} and using the fact that $P \geq 0$, we obtain
$$
v_t^\lambda + \lambda (T - t) \geq 0,
$$
which implies $P_t - v^\lambda_t \leq P_t + \lambda T$. Combined with the inequality
$$
0 \leq \frac{e^x - 1}{x} \leq e^{x^+},
$$
this yields the boundedness of $\gamma^\lambda$ under Assumption~\ref{assumption1}.

Applying It\^o's formula to the process
$e^{-\int_0^t \gamma^\lambda_s \, ds}\, v^\lambda_t$
and invoking the randomized stopping representation of optimal stopping problems in \eqref{rstopping}, we deduce that for all $\lambda >0$ and all $t\in [0,T]$
$$
v^\lambda_t \leq V_t \quad a.s.
$$

Moreover, for any $x \in \mathbb{R}$, the mapping $\lambda \mapsto \lambda\big(e^{x/\lambda} - 1\big)$
is decreasing. Indeed, a direct computation yields
\begin{gather}
\frac{d}{d\lambda}\!\Big[\lambda\big(e^{x/\lambda} - 1\big)\Big]
= \Big(1 - \frac{x}{\lambda}\Big)e^{x/\lambda} - 1 \leq 0, \label{lambdadecreasing}
\end{gather}
where the inequality follows from the elementary bound $1 + c \leq e^{c}$, valid for all $c \in \mathbb{R}$.

As a consequence, by the comparison theorem for BSDEs (see e.g. Theorem~3.4 in \cite{OZ}) and the monotone convergence theorem, there exists an optional process $
\widehat{v} := \lim_{\lambda \downarrow 0} v^\lambda$,
satisfying 
$$
\widehat{v}_t \leq V_t, \quad t \in [0,T].
$$

Our first main result shows that this inequality is, in fact, an equality.

\bt\label{T1} 
The entropy-regularized penalization scheme $v^\lambda$ converges to the value of the American option $V$ as $\lambda \rightarrow 0$, that is, it holds
$$
\lim_{\lambda \downarrow 0} v^{\lambda}_t = V_t \quad a.s.
$$
\et

\begin{proof}
To prove that $\widehat{v} = V$, it suffices to show that $\widehat{v}$ is a supermartingale dominating the payoff process $P$. Indeed, by the Snell envelope characterization of the value process $V$, this property uniquely identifies $V$ as the smallest supermartingale dominating $P$.

We begin by showing that $\widehat{v}$ is a supermartingale.  
Recall that the BSDE satisfied by $v^\lambda$ may be rewritten as
\begin{align*}
v_t^\lambda - \lambda t
= P_T - \lambda T - (m^\lambda_T - m^\lambda_t)
+ \int_t^T \lambda e^{(P_s - \lambda s - (v^\lambda_s - \lambda s))/\lambda} \, ds .
\end{align*}
Define $\bar v_t^\lambda := v_t^\lambda - \lambda t$ and $\bar P_t := P_t - \lambda t$. Then
\begin{align}
\bar v_t^\lambda
= \bar P_T - (m^\lambda_T - m^\lambda_t)
+ \int_t^T \lambda e^{(\bar P_s - \bar v^\lambda_s)/\lambda} \, ds .
\end{align}
From this representation, it follows that $(\bar v^\lambda)_{\lambda > 0}$ is a family of bounded supermartingales. Moreover, since both $v^\lambda$ and $-\lambda t$ are increasing as $\lambda \downarrow 0$, the sequence $(\bar v^\lambda)$ is increasing in $\lambda^{-1}$. Consequently, $\widehat{v} = \lim_{\lambda \downarrow 0} v^\lambda$ is the almost sure limit of an increasing sequence of supermartingales. By Theorem~1.8 in Dellacherie and Meyer \cite{DM1980}, p.~86, $\widehat{v}$ is therefore a c\`adl\`ag supermartingale.

It remains to show that $\widehat{v}$ dominates $P$.  
Since $v^\lambda \leq \widehat{v} \leq V$, we have $e^{(\widehat{v} - v^\lambda)/\lambda} \geq 1$ and therefore
\begin{align*}
\lambda \mathbb{E}[v_t^\lambda]
&= \lambda \mathbb{E}[P_T]
+ \mathbb{E}\!\left[\int_t^T \lambda^2
\Big(e^{(P_s - \widehat{v}_s)/\lambda}
e^{(\widehat{v}_s - v^\lambda_s)/\lambda} - 1\Big) ds\right] \\
&\geq \lambda \mathbb{E}[P_T]
+ \mathbb{E}\!\left[\int_t^T \lambda^2
\big(e^{(P_s - \widehat{v}_s)/\lambda} - 1\big) ds\right] \\
&\geq \lambda \mathbb{E}[P_T]
+ \mathbb{E}\!\left[\int_t^T \lambda^2
e^{(P_s - \widehat{v}_s)/\lambda}\mathbf{1}_{\{\widehat{v}_s < P_s\}} \, ds\right] - \lambda^2(T-t) \geq -\lambda^2(T-t).
\end{align*}
Applying Fatou’s lemma yields
\begin{align*}
\mathbb{E}\!\left[
\int_t^T \liminf_{\lambda \downarrow 0}
\lambda^2 e^{(P_s - \widehat{v}_s)/\lambda}
\mathbf{1}_{\{\widehat{v}_s < P_s\}} \, ds
\right] = 0 .
\end{align*}
Since $\lambda^2 e^{x/\lambda} \to +\infty$ as $\lambda \downarrow 0$ for any $x > 0$, this implies that
$\widehat{v}_s \geq P_s$ for all $s \in [t,T]$, almost surely.
Finally, as both $\widehat{v}$ and $P$ are c\`adl\`ag processes, the inequality holds for all $s \in [0,T]$.

We conclude that $\widehat{v}$ is a supermartingale dominating $P$. Since $V$ is the smallest such supermartingale, it follows that $\widehat{v} = V$.
\end{proof}

Next, we quantify the discrepancy between the entropy-regularized penalization scheme $v^\lambda$ and the value process $V$ of the American option, and derive a convergence rate in the space $\mathcal{S}^2$.

\bl \label{lemma}
For any $\lambda \in (0,1]$, the following estimate holds:
\begin{align} 
 (v^{\lambda}_t - V_t)^2 + \int_t^T d[ m^{\lambda}-M]_s 
 & \leq  2e^{T-t}\EE\!\left[\int_t^T (v^{\lambda}_s - P_s)^-\, dA_s \,\bigg|\, \cF_t\right] 
  + 2e^{T-t}\lambda^2 (T-t).
 \label{eq:penalized_eq_rate}
\end{align}
\el
\begin{proof}
Recall from \eqref{VRBSDE} that the value process admits the Doob--Meyer decomposition
$V = M - A$, where $M$ is a martingale and $A$ is a predictable, increasing process of finite variation satisfying the Skorokhod minimality condition
$$
\int_0^T (P_{s-} - V_{s-})\, dA_s = 0.
$$

For notational convenience, we denote the positive and negative parts of the driver of $v^\lambda$ by
$$
g^\lambda_s(v^\lambda)^{\pm}
:= \lambda\big(e^{(P_s - v^\lambda_s)/\lambda} - 1\big)^{\pm},
\qquad
G^{\lambda,\pm}_t(v^\lambda)
:= \int_0^t g^\lambda_s(v^\lambda)^{\pm}\, ds.
$$
Let $\beta \geq 0$. Applying It\^o’s formula to $e^{\beta t}(v^\lambda_t - V_t)^2$ and using the reflected BSDE representation \eqref{VRBSDE}, we obtain
\begin{align*}
e^{\beta t}(v^\lambda_t - V_t)^2
+ \int_t^T e^{\beta s}\, d[ N^\lambda ]_s
&= -2\int_t^T e^{\beta s}(v^\lambda_s - V_s)\, dN^\lambda_s \\
&\quad + 2\int_t^T e^{\beta s}(v^\lambda_s - V_s)\, d(G^{\lambda,+}_s - A_s) \\
&\quad - 2\int_t^T e^{\beta s}(v^\lambda_s - V_s) g^\lambda_s(v^\lambda)^-\, ds \\
&\quad - \beta\int_t^T e^{\beta s}(v^\lambda_s - V_s)^2\, ds,
\end{align*}
where we have set $N^\lambda := m^\lambda - M$.
We now focus on the reflection term. By decomposing it and using that
$\int_0^T (V_s - P_s)\, dA_s = 0$, together with the facts that $V \geq P$ and that the increasing process $G^{\lambda,+}$ grows only on the set $\{v^\lambda \leq P\}$, we obtain 
\begin{align*}
& \int^T_t e^{\beta s}(v^{\lambda}_s - V_s)  d(G^{\lambda,+} - A)_s \\
& = \int^T_t e^{\beta s}(v^{\lambda}_s - P_s)dG^{\lambda,+}_s  + \int^T_t e^{\beta s}(P_s - V_s)dG^{\lambda,+}_s  -\int^T_t e^{\beta s}(v^{\lambda}_s - P_s)dA_s  + \int^T_t e^{\beta s}(V_s- P_s)dA_s \\
& \leq e^{\beta T}\int^T_t (v^{\lambda}_s - P_s)^-  dA_s.
\end{align*}

Combining the above estimates yields
\begin{align*}
    e^{\beta t}(v^{\lambda}_t - V_t)^2  & \leq - 2\int^T_te^{\beta s}(v^{\lambda}_s - V_s) dN^{\lambda}_s + 2e^{\beta T}\int^T_t  (v^{\lambda}_s - P_s)^-  dA_s\\
    &\qquad +  2\int^T_t  e^{\beta s}(v^{\lambda}_s - V_s)[-g^\lambda_s( v^{\lambda}_s)^-]ds - \beta\int^T_t e^{\beta s}(v^{\lambda}_s - V_s)^2 ds. 
\end{align*}
Taking $\cF_t$-conditional expectations and applying Young’s inequality, we obtain
\begin{align*}
e^{\beta t}(v^{\lambda}_t - V_t)^2 & \leq  2e^{\beta T}\mathbb{E}[\int^T_t (v^{\lambda}_s - P_s)^{-}dA_s| \cF_t] + \mathbb{E}[\int^T_t e^{\beta s}( v^{\lambda}_s - V_s)^2 + e^{\beta s}[g^\lambda_s( v^{\lambda}_s)^-]^2\,\,ds | \cF_t] \\
& \qquad   - \beta\EE[\int^T_t e^{\beta s}(v^{\lambda}_s - V_s)^2 ds | \cF_{t}] - 2\EE[\int^T_te^{\beta s}(v^{\lambda}_s - V_s) dN^{\lambda}_s \, | \, \cF_{t} ].
\end{align*}
We conclude by choosing $\beta = 1$, noting that
$$
[g^\lambda_s(x)^-]^2
= \lambda^2 (e^{x/\lambda} - 1)^2 \mathbf{1}_{\{x<0\}}
\leq \lambda^2,
$$
and observing that the stochastic integral
$\left(\int_0^t e^{\beta s}(v^\lambda_s - V_s)\, dN^\lambda_s\right)_{t\geq 0}$
is a uniformly integrable martingale by the Burkholder-Davis-Gundy inequality. This yields the estimate \eqref{eq:penalized_eq_rate}.
\end{proof}

We now turn to the derivation of an explicit convergence rate of $v^\lambda$ towards $V$ as $\lambda \downarrow 0$. In view of Lemma~\ref{lemma}, it is sufficient to control the term
$$
\EE\!\left[\int_t^T (v^\lambda_s - P_s)^-\, dA_s \,\bigg|\, \cF_t\right].
$$
Obtaining sharp bounds for this quantity is a well-known difficulty in the analysis of penalisation schemes for reflected BSDEs and, in general, requires additional structural assumptions on the payoff process $P$. 

We therefore begin with an auxiliary lemma, whose proof follows the approach pioneered by El~Karoui \emph{et al.}~\cite{EKPPQ1997} and is inspired by the recent work of Gobet and Wang~\cite{GWX2023}.

\bl \label{lemma2}
There exists positive constant $C<\infty$ such that for any $\lambda \in (0,1]$ 
\begin{gather*}
\mathbb{E}\Big[\Big(\int^T_t (v^{\lambda}_s - P_s)^{-}ds\Big)^2 \Big| \cF_t\Big] \leq C(\lambda -\lambda \ln \lambda)^2.
\end{gather*}

\el 
\begin{proof}
By applying the mean value theorem to the function $x \mapsto \lambda\big(e^{x/\lambda}-1\big)$ at the point $x=\varepsilon>0$, we obtain the following lower bound for the driver:
\begin{align*}
\lambda(e^{x/\lambda} - 1) 
& \geq \lambda(e^{(x\wedge \epsilon)/\lambda} - 1)  +  (x-\epsilon)^+e^{\epsilon/\lambda}\\
& \geq \lambda(e^{(x\wedge \epsilon)/\lambda} - 1) - \epsilon e^{\epsilon/\lambda} +  x^+e^{\epsilon/\lambda},
\end{align*}
where, in the last inequality, we have used the elementary bound $(x-\varepsilon)^+ \ge x^+ - \varepsilon$.

Motivated by this estimate, and for fixed $\lambda\le 1$, we introduce the following BSDE with a Lipschitz continuous driver:
\begin{align}
    X^\lambda_{t}
   & = P_{T} - (M^X_{T} - M^{X}_{t}) + \int_{t}^{T} f_{\lambda,\epsilon}(P_s-X^\lambda_s) ds + \int_{t}^{T}  (P_s  - X^\lambda_s )^+\gamma^{\lambda,\epsilon} ds, \label{Xpen}
\end{align}
where $f_{\lambda, \epsilon}(x) := \lambda(e^{(x\wedge \epsilon)/\lambda} -1 )-\epsilon e^{\epsilon/\lambda}$  and  $\gamma^{\lambda, \epsilon} := e^{\epsilon/\lambda}$. By the comparison theorem for BSDEs (see Theorem~3.4 in \cite{OZ}), it then follows that for all $t\in [0,T]$ and all $\lambda\in (0,1]$
\begin{equation}\label{two:sided:bounds:vlambda}
X^\lambda_t \le v^\lambda_t \le V_t, \quad a.s.
\end{equation}

By arguments similar to those used in Lemma~\ref{lemma}, we obtain
\begin{align*}
(X^\lambda_t)^2 + \int^T_t d[M^X]_s
&= P^2_T - \int^T_t 2X^\lambda_s dM^X_s+ \int^T_t 2X_s f_{\lambda,\epsilon}(P_s-X^\lambda_s)ds + \int^T_t 2X^\lambda_s (P_s -X^\lambda_s)^+ \gamma^{\lambda,\epsilon} ds \\
        & \leq P^2_T - \int^T_t 2X^\lambda_s dM^{X}_s+ \int^T_t 2X^\lambda_s f_{\lambda,\epsilon}(P_s-X^\lambda_s)ds + \int^T_t 2P_s (P_s -X^\lambda_s)^+ \gamma^{\lambda,\epsilon} ds\\
                & \leq P^2_T - \int^T_t 2X^{\lambda}_s dM^{X}_s + \int^T_t (X^{\lambda}_s)^{2} ds + \int^T_t (f_{\lambda,\epsilon}(P_s-X^\lambda_s))^2 ds + \frac{2}{\alpha} (\sup_{0\leq s\leq T}P_s)^2 \\ 
                &\quad + 2\alpha\Big(\int^T_t  (P_s -X^\lambda_s)^+ \gamma^{\lambda,\epsilon} ds\Big)^2.
\end{align*}
where the last inequality follows from Young’s inequality, for an arbitrary $\alpha>0$.

Applying Jensen’s and Cauchy–Schwarz inequalities yields
\begin{equation}\label{ineq:positive:part:squared}
\begin{aligned}
\EE\Big[\Big(\int^T_t  (P_s -X^\lambda_s)^+ \gamma^{\lambda,\epsilon} ds\Big)^2\Big] 
& \leq 4\EE[(X^\lambda_t)^2] + 4\EE[P_T^2] + 4\EE[[M^X]_T-[M^X]_t] \\
& \qquad + 4T\int^T_t \EE[(f_{\lambda,\epsilon}(P_s-X^\lambda_s))^2] ds.
\end{aligned}
\end{equation}
Choosing $\alpha$ such that $2\alpha = \tfrac{1}{12}$, and using the bound 
\begin{equation}\label{bound:f:lambda:epsilon}
(f_{\lambda, \epsilon}(x))^2 \leq 2\lambda^2(e^{(x\wedge \epsilon)/\lambda} -1 )^2 + 2\epsilon^2 e^{2\epsilon/\lambda} \leq 2\lambda^2e^{2\epsilon/\lambda}  + 2\epsilon^2 e^{2\epsilon/\lambda},
\end{equation}
we deduce
\begin{align*}
\frac{2}{3}\EE[(X^\lambda_t)^2] + \frac{2}{3}\EE\Big[\int^T_t d[M^X]_t\Big]
                & \leq C\left(1 + \int^T_t \EE[(X^\lambda_s)^2] ds + \int^T_t \EE[(f_{\lambda,\epsilon}(P_s-X^\lambda_s))^2] ds\right)\\
                & \leq C\left(1 + (\lambda^2 + \epsilon^2) e^{2\epsilon/\lambda} + \int^T_t \EE[(X^\lambda_s)^2] ds  \right).
\end{align*}

An application of Grönwall’s lemma then yields
\begin{equation}\label{bound:l2:moment:x}
\sup_{0\leq t\leq T} \EE[|X^\lambda_t|^2] \leq  C\big(1 + (\lambda^2 + \epsilon^2)e^{2\epsilon/\lambda}\big),
\end{equation}
and consequently, 
\begin{equation}\label{bound:quadratic:variation:M}
\EE[[M^X]_T] \leq  C\big(1 + (\lambda^2 + \epsilon^2)e^{2\epsilon/\lambda}\big).
\end{equation}

Combining \eqref{two:sided:bounds:vlambda} with \eqref{ineq:positive:part:squared}, \eqref{bound:f:lambda:epsilon}, \eqref{bound:l2:moment:x} and \eqref{bound:quadratic:variation:M}, we get
\begin{align}
\mathbb{E}\Big[\Big(\int^T_0 (v^{\lambda}_s - P_s)^{-}ds\Big)^2\Big]
 \leq \EE\Big[\Big(\int_{0}^{T}  (X^\lambda_s - P_s)^-ds\Big)^2\Big] 
 \leq C(e^{-2\epsilon/\lambda} + \lambda^2 + \epsilon^2). \label{prate}
\end{align}

Next, we choose $\epsilon$ so as to optimize the convergence rate. Observe first that the constraint
$\epsilon/\lambda \geq 1$ is necessary, since this quantity must diverge as $\lambda \to 0$.
Consequently, the two dominant (slowest-decaying) terms are $e^{-2\epsilon/\lambda}$ and $\epsilon^2$.
Balancing these contributions leads us to impose $\epsilon = e^{-\epsilon/\lambda}$
or, equivalently, $\ln \epsilon = -\epsilon/\lambda$. This transcendental equation does not admit a closed-form solution, and we therefore seek an accurate approximation.

Rewriting the above identity yields
\begin{align*}
 -\frac{\epsilon}{\lambda} = \ln (\lambda) + \ln \frac{\epsilon}{\lambda} - \ln 1.
\end{align*}
By the mean value theorem, there exists some  $c \in [1, \epsilon/\lambda]$,
\begin{align*}
\epsilon & = -\lambda \frac{1}{c}(\frac{\epsilon}{\lambda}-1) - \lambda \ln (\lambda) = -\frac{1}{c}(\epsilon-\lambda) - \lambda \ln (\lambda)
\end{align*}
\noindent which gives
\begin{align}
    \epsilon =\left(\frac{1}{1+1/c}\right) \left(\frac{\lambda}{c} - \lambda\ln \lambda\right) \leq \lambda- \lambda \ln \lambda. \label{opepsilon}
\end{align}
Motivated by this estimate, we adopt the approximate choice $\epsilon = \lambda - \lambda \ln \lambda$ which clearly satisfies $\epsilon \geq \lambda$ for all $\lambda \in (0,1]$.
Substituting this value into~\eqref{prate} readily gives the desired upper-bound.
\end{proof}

As in the classical penalization approach, deriving the convergence rate towards the value $V$ of the American option with respect to the temperature parameter $\lambda$ requires additional
regularity assumptions, which we now state.

\bhyp \label{A2}
The filtration $\FF$ is generated by a Brownian motion, and the payoff process $P$ admits the
generalized semimartingale decomposition
$$
P_{t}
= P_{0}
+ \int_{0}^{t} U_{s}\,ds
+ \int_{0}^{t} V_{s}\,dW_{s}
+ H_{t},
$$
\noindent where $U, V \in \cS^{2}$, and $H$ is a continuous, non-decreasing process satisfying
$H_{0}=0$ and $H_{T}\in L^{2}$.
\ehyp

Under Assumption~\ref{A2}, it follows from Proposition~4.2 and Remark~4.3 in~\cite{EKPPQ1997}
that the reflection process $A$ is absolutely continuous with respect to Lebesgue measure.
Moreover, its density is uniformly dominated, in the sense that $dA_{t} \leq \kappa_{t}\,dt$, with $\kappa := U^{-}$.
As an illustration, the standard American put option satisfies Assumption~\ref{A2};
see Remark~2.3 in~\cite{GWX2023}.

We are now in a position to state the main convergence result with respect to the temperature
parameter $\lambda$. Under the additional regularity imposed by Assumption~\ref{A2}, the
penalised value process converges to the American option value at an explicit rate in the
$\cS^{2}$-norm.

\bt \label{theorem: s2_convergence}
Under Assumption~\ref{A2}, there exists a positive constant $C$ such that, for any $\lambda \in\! (0,1]$,
    \begin{equation*}
        \|v^{\lambda} - V\|_{\cS^2} \leq C(\lambda -\lambda \ln \lambda).
    \end{equation*}
\et
\begin{proof}
The proof relies on the absolute continuity of the reflection process ensured by
Assumption~\ref{A2}. More precisely, by Proposition~4.2 and Remark~4.3 in
El~Karoui et al.~\cite{EKPPQ1997}, the increasing process $A$ satisfies
$dA_t \leq \kappa_t\,dt$, where $\kappa_t \leq U_t^{-}$. It follows that
\begin{align*}
\EE[ \int^T_0 (Y^\lambda_s - P_s)^- dA_s ] & \leq \EE[\sup_{0\leq t\leq T} \kappa_t \int^T_0 (Y^\lambda_s - P_s)^- ds]\\
& \leq \EE[\big(\sup_{0\leq t\leq T} \kappa_t\big)^2 ]^\frac{1}{2}\EE[|\int^T_0 (Y^\lambda_s - P_s)^- ds\big|^2]^\frac{1}{2}.
\end{align*}
where the second inequality follows from the Cauchy-Schwarz inequality.
The conclusion then follows by combining Lemma~\ref{lemma} and Lemma~\ref{lemma2}.
\end{proof}

As indicated by the two-sided bounds in~\eqref{two:sided:bounds:vlambda}, the process $v^\lambda$ provides a lower bound for the American option value $V$. We now complement this result by deriving a convergence rate for a corresponding upper bound, based on the dual representation of Rogers~\cite{R2002}. To this end, we introduce the process $u^\lambda$ defined by
$$
u^\lambda_t
:= \EE\!\left[\sup_{t \leq \sigma \leq T} \bigl(P_\sigma - m^\lambda_\sigma\bigr)
\,\big|\, \cF_t \right] + m^\lambda_t,
$$

\noindent where $m^\lambda$ denotes the martingale component arising in the entropy-regularized penalization scheme~\eqref{dongbsdemod}. Our goal is to quantify the rate at which $u^\lambda$ converges to $V$.

\bt \label{theorem: Us2_convergence}
Under Assumption~\ref{A2}, there exists a constant $C<\infty$ such that, for any $\lambda \in (0,1]$,
$$
\|u^\lambda - V\|_{\cS^2}
\leq C\bigl(\lambda - \lambda \ln \lambda\bigr).
$$
\et
\begin{proof}
The argument relies on the fact that $V = M - A$ dominates the payoff process $P$, together with
the monotonicity of the reflection process $A$. Indeed, for any $t \in [0,T]$, Theorem 2.1 of Rogers~\cite{R2002} gives $u^{\lambda}_{t} \geq V_{t}$, and we have
\begin{align*}
0 \leq u^\lambda_t - V_t
&= \EE\!\left[\sup_{t \leq \sigma \leq T} (P_\sigma - m^\lambda_\sigma)
\,\big|\, \cF_t \right]
+ m^\lambda_t - V_t \\
&\leq \EE\!\left[\sup_{t \leq \sigma \leq T} (M_\sigma - m^\lambda_\sigma)
\,\big|\, \cF_t \right]
- A_t + m^\lambda_t - V_t \\
&\leq \EE\!\left[\sup_{0\leq s \leq T} |M_s - m^\lambda_s|
\,\big|\, \cF_t \right]
+ |m^\lambda_t - M_t|.
\end{align*}
For notational convenience, we set $N^\lambda := m^\lambda - M$ and define
$$
n^\lambda_t := \EE\!\left[\sup_{0\leq s \leq T} |N^\lambda_s| \,\big|\, \cF_t \right].
$$

Applying Jensen's inequality yields
\begin{align*}
(u^\lambda_t - V_t)^2
&\leq 2|n^\lambda_t|^2 + 2|N^\lambda_t|^2 
\leq 2\Big(\sup_{0\leq t \leq T} |n^\lambda_t|\Big)^2
+ 2\Big(\sup_{0\leq t \leq T} |N^\lambda_t|\Big)^2.
\end{align*}
Using repeated applications of the Burkholder-Davis-Gundy inequality together with
It\^o's isometry, we obtain
\begin{align*}
    \EE[\sup_{0\leq t\leq T} (u^\lambda_t - V_t)^2]
        & \leq 2\EE[|\sup_{0\leq t\leq T}|n^\lambda_t||^2]  + 2\EE[|\sup_{0\leq t\leq T}|N^\lambda_t||^2].\\
        & \leq 2C\EE[[n^\lambda]_T]  + 2C\EE[[N^\lambda]_T].\\
        & = 2C\EE[|n^\lambda_T|^2]  + 2C\EE[[N^\lambda]_T]\\
        & \leq 2C\EE[|\sup_{0\leq t\leq T} |N^\lambda_t||^2]  + 2C\EE[[N^\lambda]_T] \leq  4C\EE[[m^\lambda-M]_T].
\end{align*}
The conclusion follows from Lemma~\ref{lemma} and Lemma~\ref{lemma2}.
\end{proof}

\subsection{Policy Improvement Algorithm}
In this section, we introduce the PIA associated with the
entropy-regularized formulation and analyze its convergence properties. The proposed approach
is inspired by the methodology developed in~\cite{CFL2026}, and relies on an iterative procedure
alternating between policy evaluation and policy improvement. At each iteration, the policy
evaluation step consists in solving a linear BSDE,
while the policy improvement step is obtained by maximizing, in closed form, the corresponding
regularized Hamiltonian.

To this end, for a fixed temperature parameter $\lambda>0$, we define the function
\begin{align*}
G(s,x,\pi_s)
&:= (P_s - x)\pi_s - \lambda\bigl(\pi_s \ln \pi_s - \pi_s + 1\bigr),
\end{align*}
and denote by
$$
\pi_s^{*}(x)
:= \arg\max_{\pi} G(s,x,\pi_s)
= \exp\!\left(\frac{P_s - x}{\lambda}\right).
$$
Let $\pi^0 \in \Pi$ be an arbitrary initial policy, and define the corresponding initial value
function by $v^{\lambda,0}_t := \EE[P_T \mid \cF_t]$ for $t \in [0,T]$. Given the current iterate
$(\pi^m, v^{\lambda,m}) \in \Pi \times \mathcal{D}$, the $(m+1)$-th policy improvement step is
constructed as follows. The policy is first updated according to
\begin{equation}
\pi^{m+1}_s
:= \pi^{*}_s\!\left(v^{\lambda,m}_s\right)
= \exp\!\left(\frac{P_s - v^{\lambda,m}_s}{\lambda}\right),
\label{eq: pin1}
\end{equation}
and the corresponding value function $v^{\lambda,m+1}$ is then obtained as the unique solution
to the linear BSDE
\begin{equation}
v^{\lambda,m+1}_t
= P_T - \bigl(m^{\lambda,m+1}_T - m^{\lambda,m+1}_t\bigr)
+ \int_t^T G\bigl(s, v^{\lambda,m+1}_s, \pi^{m+1}_s\bigr)\,ds, \quad t \in [0,T].
\label{eq: b_Vlamb_n}
\end{equation}

We note that, for each $m \geq 0$, the process $v^{\lambda,m}$ is well defined and belongs to
$\mathcal{D}$. Moreover, since by construction,
$$
G\bigl(s, v^{\lambda,m}_s, \pi^{m}_s\bigr)
\leq
G\bigl(s, v^{\lambda,m}_s, \pi^{m+1}_s\bigr) \quad a.s.,
$$
the comparison theorem for BSDEs (see e.g. Theorem~3.4 in~\cite{OZ}) implies that the sequence
$\{v^{\lambda,m}\}_{m\geq0}$ is non-decreasing. That is, for all integer $m$ and all $t\in[0,T]$
$$
v^{\lambda,m}_t \leq v^{\lambda,m+1}_t \quad a.s..
$$

We now establish the convergence rate of the above policy improvement scheme. The following result
shows that, for a fixed temperature parameter $\lambda$, the sequence of value functions
$\{v^{\lambda,m}\}_{m\geq0}$ converges monotonically to the entropy-regularized value function
$v^{\lambda}$ at a factorial rate.

\bt  \label{thm: policy_convergence2}
There exists $C<\infty$ such that for any $\lambda \in(0,1]$ and any $t \in [0,T)$, 
\begin{gather*}
0\leq v_{t}^{\lambda} - v_{t}^{\lambda,m} \leq \frac{(CT)^m}{m!} \mathbb{E}\Big[\sup_{0\leq s\leq T} (v^{\lambda, 1}_s - v^{\lambda, 0}_s)\,\Big|\, \cF_t\Big].
\end{gather*}
\et 
\begin{proof}
The driver at step $m+1$ is given by
\begin{align} \label{eq: Gpi_def1}
    G(s, v_{s}^{\lambda, m+1}, \pi_{s}^{m+1}) & = (P_{s} - v_{s}^{\lambda, m+1})\pi^{m+1}_s - \lambda(\pi^{m+1}_s \ln(\pi^{m+1}_s) - \pi^{m+1}_s  + 1)\nonumber \\
    & = e^{\frac{P_s - v^{\lambda,m}_s}{\lambda}} (v_{s}^{\lambda,m} - v_{s}^{\lambda, m+1} ) + \lambda (e^{\frac{P_s - v^{\lambda,m}_s}{\lambda}} - 1).
\end{align}
Moreover, by Assumption \ref{assumption1} the payoff $P$ is non-negative and bounded and therefore the initial value $v^{\lambda,0}$ is also bounded. Hence, by Theorem~3.3 in~\cite{OZ} we deduce that $v^{\lambda,1}$ is non-negative and thus $v^{\lambda, m} \geq 0$ for all $m$. In view of this, the difference of the drivers can be estimated as
\begin{align*}
    &G(s, v_{s}^{\lambda, m+1}, \pi_{s}^{m+1}) -  G(s, v_{s}^{\lambda, m}, \pi_{s}^{m}) \\
    &= e^{\frac{P_s - v^{\lambda,m}_s}{\lambda}} (v_{s}^{\lambda,m} - v_{s}^{\lambda, m+1} ) - e^{\frac{P_s - v^{\lambda,m-1}_s}{\lambda}} (v^{\lambda,m-1}_s - v^{\lambda, m}_s ) + \lambda (e^{\frac{P_s - v^{\lambda,m}_s}{\lambda}} - e^{\frac{P_s - v^{\lambda,m-1}_s}{\lambda}}) \\
    &\leq  e^{\frac{P_s - v^{\lambda,m-1}_s}{\lambda}} (v^{\lambda,m}_s - v^{\lambda, m-1}_s ) \\
    &\leq C (v^{\lambda,m}_s - v^{\lambda, m-1}_s ),
\end{align*}
where $C$ is a constant independent of $m$. By the previous inequality and the Fubini-Tonelli theorem, we have
\begin{align*}
0\leq v^{\lambda, m+1}_t - v^{\lambda, m}_t & = \mathbb{E}[\int^T_t \Big( G(s, v_{s}^{\lambda, m+1}, \pi_{s}^{m+1}) -  G(s, v_{s}^{\lambda, m}, \pi_{s}^{m}) \Big) \, ds   \,|\, \cF_t]\\
& \leq C\mathbb{E}[\int^T_t (v^{\lambda, m}_{t_{m-1}} - v^{\lambda, m-1}_{t_{m-1}}) \, d{t_{m-1}}   \,|\, \cF_t]\\
& \leq C^m\mathbb{E}[\int^T_{t}\int^T_{t_{m-1}}\dots \int^T_{t_1} (v^{\lambda, 1}_{t_0} - v^{\lambda, 0}_{t_0})\, dt_0\dots dt_{m-2}dt_{m-1}  \,|\, \cF_t]\\
& \leq \frac{(CT)^m}{m!} \mathbb{E}\big[\sup_{0\leq s\leq T} (v^{\lambda, 1}_{s} - v^{\lambda, 0}_s)\,\big|\, \cF_t\big].
\end{align*}

On the other hand, by the mean value theorem $G(s, v_{s}^{\lambda, m+1}, \pi_{s}^{m+1}) \leq \lambda e^{\frac{P_{s} - v_{s}^{\lambda, m+1}}{\lambda}}$ and by similar computations we have
\begin{align*}
    0 \leq v^{\lambda}_{t} - v^{\lambda, m+1}_{t} & = \mathbb{E}\Big[\int^T_t \Big( \lambda(e^{\frac{P_{s} - v_{s}^{\lambda}}{\lambda}} - 1) -  G(s, v_{s}^{\lambda, m+1}, \pi_{s}^{m+1}) \Big) \, ds   \,|\, \cF_t \Big] \\
    & = \mathbb{E}\Big[\int^T_t \Big( \lambda(e^{\frac{P_{s} - v_{s}^{\lambda}}{\lambda}} - e^{\frac{P_s - v^{\lambda,m}_s}{\lambda}}) + e^{\frac{P_s - v^{\lambda,m}_s}{\lambda}} (v_{s}^{\lambda, m+1} - v_{s}^{\lambda,m}) \Big) \, ds   \,|\, \cF_t \Big] \\
    &\leq C\mathbb{E}\Big[\int^T_t (v_{s}^{\lambda, m+1} - v_{s}^{\lambda,m} ) \, ds   \,|\, \cF_t\Big] \\
    & \leq \frac{(CT)^{m+1}}{(m+1)!} \mathbb{E}\big[\sup_{0\leq s\leq T} (v^{\lambda, 1}_{s} - v^{\lambda, 0}_s)\,\big|\, \cF_t\big].
\end{align*}
This completes the proof. 
\end{proof}

\section{Numerical Experiments}\label{sec:numerics}

We now turn to numerical experiments in order to illustrate the performance of the proposed iterative scheme and to assess its accuracy in practical settings.

Starting from an initial guess $v^{\lambda,0}>0$, we observe that, in the presence of a non-zero interest rate, the sequence $\{v^{\lambda,m}\}_{m\geq 0}$ satisfies at each iteration a linear BSDE of the form
\begin{align}\label{dynamics:vlambda:m:recall}
v^{\lambda,m+1}_t
&= P_T - (m^{\lambda,m+1}_T - m^{\lambda,m+1}_t)
 + \int_{t}^{T} \Big( G(s, v^{\lambda,m+1}_{s}, \pi^{m+1}_{s}) - r v^{\lambda,m+1}_s \Big) ds,
\end{align}
where the function $G$ is defined in \eqref{eq: Gpi_def1} and $r>0$ denotes the constant interest rate. It follows from \eqref{dynamics:vlambda:m:recall} that, for any $\Delta t \ge 0$, the process $v^{\lambda,m+1}t$ satisfies
\begin{align*}
v^{\lambda, m+1}_{t}
    &=  \mathbb{E}\Big[ e^{-\int^{t + \Delta t}_t a^{m}_{s} ds}  v^{\lambda,m+1}_{t+ \Delta t} + \int_{t}^{t+\Delta t} e^{-\int_{t}^s a^{m}_u \, du} \, b^{m}_s \, ds  \,\Big|\, \mathcal{F}_{t} \Big].
\end{align*}
For small $\Delta t$, a first-order approximation over the interval $[t,t+\Delta t]$ then yields
\begin{align*}
 v^{\lambda, m+1}_{t}   & \approx e^{-a_{t}^{m}\Delta t} \EE\left[v^{\lambda,m+1}_{t+ \Delta t} \Big | \cF_{t}\right] + \frac{b^{m}_{t}}{a^{m}_{t}}(1 - e^{-a^{m}_{t}\Delta t}).
\end{align*}
Here, the processes $a^{m}$ and $b^{m}$ are defined for $t\in[0,T]$ by
$$
a^{m}_t = e^{\frac{P_t - v^{\lambda, m}_t}{\lambda}} + r, \qquad
b^{m}_t = e^{\frac{P_s - v^{\lambda, m}_t}{\lambda}} \, v^{\lambda, m}_s
      + \lambda \Big( e^{\frac{P_t - v^{\lambda, m}_t}{\lambda}} - 1 \Big), \quad t\in [0,T].
$$
An important feature of this formulation is that $v^{\lambda,m+1}$ can be computed explicitly in terms of $v^{\lambda,m}$.

The above discussion suggests the following iterative scheme. For a given positive integer $N$, we let $\Delta t= T/N$ and define the uniform time grid $t_{k}=k \Delta t$, $k=0,\cdots, N$. We then let $\bar{v}^{\lambda, 0}_{t_k}   = \mathbb{E}[e^{-r(T-t)}P_T| \mathcal{F}_{t_k}]$, $k=0,\cdots, N$, which corresponds to the price of the European option. Then, for any integer $m$ and any $k=N-1, \cdots, 0$, 
\begin{align*}
\bar{v}^{\lambda, m+1}_{t_k}
& = e^{-\bar{a}_{t_k}^{m}\Delta t} \EE\left[\bar{v}^{\lambda,m+1}_{t_{k+1}} \Big | \cF_{t_k}\right] + \frac{\bar{b}^{m}_{t_k}}{\bar{a}^{m}_{t_k}}(1 - e^{-\bar{a}^{m}_{t_k}\Delta t}),
\end{align*}

\noindent where the process $\bar{a}^m$ and $\bar{b}^m$ are obtained from $a^m$ and $b^m$ by replacing $v^{\lambda, m}$ by its approximation $\bar{v}^{\lambda, m}$. The conditional expectation can be estimated using least squares regression. In the example below we regress on the 13 basis functions suggested by Andersen and Broadie \cite{AB2004}. 

We test the above scheme on the symmetric case of an American max-call option. To be specific, given the strike price $K$, we recall that the price at time $0$ of a max-call option is defined by
\begin{equation*}
    \sup_{ \tau \in \mathcal{T}_{0,T}} \mathbb{E}\!\left[e^{-r\tau} \Big(\max_{1 \leq i \leq d} S_{\tau}^{i} - K\Big)^{+}\right].
\end{equation*}
We assume that the underlying assets follow a $d$-dimensional Black–Scholes model with dividends:
\begin{equation} \label{eq: GBMwD}
S_t^i = s_0^i \exp\big((r - \delta - \sigma^2/2) t + \sigma W_t^i\big), \quad i = 1, \dots, d,
\end{equation}

\noindent where $s_0^i$ denotes the initial values, $r$ the risk-free rate, $\delta$ the constant dividend yield, $\sigma$ the constant volatility, and $W=(W^1, \cdots, W^d)$ a standard $d$-dimensional Brownian motion. 

Below, we present our results and compare them to prices computed using the classical penalization approach of El Karoui \emph{et al.} \cite{EKPPQ1997} and a binomial tree approximation. 

Table \ref{table1} indicates that the PIA scheme delivers stable and accurate approximations of the American max-call price. As the parameter $\lambda$ decreases, the PIA values converge monotonically and become very close to those obtained with the classical penalization or binomial approaches, thereby confirming the consistency of the method. For $\lambda = 0.001$, the discrepancy between PIA and classical penalization is negligible across all tested initial prices. Moreover, the resulting prices are in good agreement with the binomial benchmark, with deviations remaining small and comparable to those observed for the classical penalization scheme. Overall, these results suggest that the proposed PIA provides a reliable and numerically efficient procedure for the valuation of American-style derivatives.

\begin{table}[h!]
\centering
\begin{tabular}{r r | c |>{\centering\arraybackslash}m{2.5cm} >{\centering\arraybackslash}m{2.5cm}}
{$S_0$} & {$\lambda$} & {PIA} & {Classical penalization} & {Binomial} \\
\hline
\hline
90 & 0.1 & 8.063 & 8.208 & 8.296 \\
90 & 0.01 & 8.380 & 8.424 & 8.296 \\
90 & 0.001 & 8.428 & 8.460 & 8.296 \\
\hline
100 & 0.1 & 14.017 & 14.040 & 14.211 \\
100 & 0.01 & 14.362 & 14.357 & 14.211 \\
100 & 0.001 & 14.412 & 14.408 & 14.211 \\
\hline
110 & 0.1 & 21.613 & 21.494 & 21.799 \\
110 & 0.01 & 21.971 & 21.914 & 21.799 \\
110 & 0.001 & 22.021 & 21.980 & 21.799 \\
\hline
\end{tabular}
\caption{Price at time $0$ of the American max-call option with parameters: $d=2$, $s^1_0=s^2_0=s_0$, $K = 100$, $r = 0.05$, $\sigma = 0.2$, $\delta = 0.1$, $T = 3$ and $N=100$ and $100000$ sample paths. The PIA is computed over 2000 iterations.}
\label{table1}
\end{table}

\brem
In the above implementation, we notice that for $v_{s}^{\lambda, m} < P_{s}$ we have
\begin{equation*}
    v^{\lambda, m+1}_{t} \approx \frac{b^{m}_{t}}{a^{m}_{t}}(1 - e^{-a^{m}_{t}\Delta t}) \approx v_{t}^{\lambda, m} + \lambda.
\end{equation*}
Namely, the optimal policy (or stopping intensity), which is of the exponential form given by 
$\exp\!\left(\frac{P_s - v^{\lambda,m}_s}{\lambda}\right),$
can significantly impact the update speed and the convergence of the value function when $\lambda$ is small and $v^{\lambda, m} < P$. Consequently, for small $\lambda$, the choice of initialization can heavily influence the computational time required for the scheme to converge. For the results presented in Table \ref{table1}, we observe this behaviour for the cases of $\lambda = 0.01$ and $\lambda = 0.001$. 

To overcome this, we consider scheduling $\lambda$ by first employing large $\lambda$ to prioritize convergence speed, and subsequently transition to a small $\lambda$ to ensure accuracy. Namely, for the case of $\lambda = 0.001$ we sequentially run the PIA for $\lambda =0.1, 0.05, 0.01$ and finally $\lambda = 0.001$ for $500$ iterations each.
\erem


\vskip-15pt\vskip-15pt

\begin{thebibliography}{99} {\parskip = 1pt

\bibitem{AB2004}
Andersen, L., \& Broadie, M. (2004). Primal-dual simulation algorithm for pricing multidimensional American options. \textit{Management Science, 50}(9), 1222--1234.


\bibitem{BCJ2019}
Becker, S., Cheridito, P., \& Jentzen, A. (2019). Deep optimal stopping. \textit{Journal of Machine Learning Research, 20}, 74.

\bibitem{BCJW2021}
Becker, S., Cheridito, P., Jentzen, A., \& Welti, T. (2021). Solving high-dimensional optimal stopping problems using deep learning. \textit{European Journal of Applied Mathematics, 32}, 470-514.

\bibitem{BCJ2020}
Becker, S., Cheridito, P., \& Jentzen, A. (2020). Pricing and hedging American-style options with deep learning. \textit{Journal of Risk and Financial Management, 13}(7), 158.

\bibitem{BG2004}
Broadie, M., \& Glasserman, P. (2004). A stochastic mesh method for pricing high dimensional American options. \textit{Journal of Computational Finance, 7}, 35-72.


\bibitem{CFL2025}
Chee, D., Frikha, N., \& Li, L. (2025). An entropy regularized BSDE approach to Bermudan options and games. \textit{arXiv preprint arXiv:2509.18747}. \url{https://arxiv.org/abs/2509.18747}

\bibitem{CFL2026}
Chee, D., Frikha, N., \& Li, L. (2026). Entropy-regularized penalization schemes for American options and reflected BSDEs with singular generators. \textit{Working Paper}. 


\bibitem{DM1980}
Dellacherie, C., \& Meyer, P. A. (1980). \textit{Probabilités et potentiel. Chap. V-VIII}. Hermann.


\bibitem{DDJZ2023}
Dai, M., Dong, Y., Jia, Y., \& Zhou, X. (2023). Learning Merton’s strategies in an incomplete market: Recursive entropy regularization and biased Gaussian exploration. arXiv:2312.11797 [math.OC]. Retrieved from https://arxiv.org/abs/2312.11797

\bibitem{DD2024}
Dai, M., \& Dong, Y. (2024). Learning an optimal investment policy with transaction costs via a randomized Dynkin game. SSRN Working Paper. Retrieved from https://ssrn.com/abstract=4871712

\bibitem{DSXZ2024}
Dai, M., Sun, Y., Xu, Z. Q., \& Zhou, X. Y. (2024). Learning to optimally stop a diffusion process. \textit{arXiv preprint arXiv:2408.09242}. \url{https://arxiv.org/abs/2408.09242}

\bibitem{DD2024}
Dai, M., \& Dong, Y. (2024). Learning an optimal investment policy with transaction costs via a randomized Dynkin game. \textit{SSRN}. \url{https://ssrn.com/abstract=4871712}

\bibitem{D2023}
Dong, Y. (2024). Randomized optimal stopping problem in continuous time and reinforcement learning algorithm. \textit{SIAM Journal on Control and Optimization, 62}(3), 1590--1614.

\bibitem{DZ2025}
Dong, Y., \& Zheng, H. (2025). Extended HJB equation for mean–variance stopping problem: Vanishing regularization method. \textit{arXiv preprint arXiv:2510.24128}, \url{https://arxiv.org/abs/2510.24128}


\bibitem{DFX2024}
Dianetti, J., Ferrari, G., \& Xu, R. (2024). Exploratory optimal stopping: A singular control formulation. \textit{arXiv preprint arXiv:2408.09335}. \url{https://arxiv.org/abs/2408.09335}

\bibitem{EKPPQ1997}
El Karoui, N., Kapoudjian, C., Pardoux, E., Peng, S., \& Quenez, M. C. (1997). Reflected solutions of backward SDEs, and related obstacle problems for PDEs. \textit{Annals of Probability, 25}(2), 702-737.



\bibitem{GWX2023}
Gobet, E., \& Wang, W. (2026). Improved Convergence Rate for Reflected BSDEs by Penalization Method. \textit{Applied Mathematics \& Optimization}, 93(10).

\bibitem{GIOOQ2017}
Grigorova, M., Imkeller, P., Offen, E., Ouknine, Y., \& Quenez, M. C. (2017). Reflected BSDEs when the obstacle is not right-continuous and optimal stopping. \textit{Ann. Appl. Probab}, 27(5), 3153-3188.


\bibitem{GS2008}
Gyöngy, I., \& Šiška, D. (2008). On randomized stopping. \textit{Bernoulli, 14}(2), 352--361.

\bibitem{HO2015}
Hamadène, S., \& Ouknine, Y. (2016). Reflected backward SDEs with general jumps. \textit{Theory of Probability \& Its Applications, 60}(2), 263--280.

\bibitem{HLYZ2025}
Huang, Y., Li, M., Yu, X., \& Zhou, Z. (2025). Continuous-time reinforcement learning for optimal switching over multiple regimes. arXiv:2512.04697 \url{https://arxiv.org/abs/2512.04697v2}

\bibitem{JWW2024}
Jia, B., Wang, L., \& Wong, H. Y. (2024). Machine learning of surrender: Optimality and humanity. Journal of Risk and Insurance, 91(4), 915–942.





\bibitem{LMX2005}
Lepeltier, J. P., Matoussi, A., \& Xu, M. (2005). Reflected backward stochastic differential equations under monotonicity and general increasing growth conditions. \textit{Advances in Applied Probability, 37}, 134--159.


\bibitem{LS2021}
Longstaff, F. A., \& Schwartz, E. S. (2001). Valuing American options by simulation: A simple least-squares approach. \textit{Review of Financial Studies, 14}(1), 113--147.


\bibitem{M1978}
Maingueneau, A. M. (1978). Temps d'arrêt optimaux et théorie générale. In C. Dellacherie, P. A. Meyer, \& M. Weil (Eds.), \textit{Séminaire de Probabilités XII, Lecture Notes in Mathematics} (Vol. 649, pp. 457--467). Springer.


\bibitem{OZ}
Øksendal, B., \& Zhang, T. (2012). Backward stochastic differential equations with respect to general filtrations and applications to insider finance. \textit{Communications on Stochastic Analysis, 6}(4), Article 13.

\bibitem{P1999}
Peng, S. (1999). Monotonic limit theorem of BSDE and nonlinear decomposition theorem of Doob–Meyers type. \textit{Probability Theory and Related Fields, 113}, 473-499.


\bibitem{STG2023}
Soner, H. M., \& Tissot-Daguette, V. (2025). Stopping times of boundaries: Relaxation and continuity. \textit{SIAM Journal on Control and Optimization, 63}(4), 2835--2855. 

\bibitem{RST2022}
Reppen, A. M., Soner, H. M., \& Tissot-Daguette, V. (2025). Neural optimal stopping boundary. {\it Mathematical Finance}, 35, 441–469. https://doi.org/10.1111/mafi.12450

\bibitem{R2002}
Rogers, L. C. G. (2002). Monte Carlo valuation of American options. \textit{Mathematical Finance, 12}(3), 271-286.

\bibitem{TZZ2021}
Tang, W., Zhang, P. Y., \& Zhou, X. Y. (2023). Exploratory HJB equations and their convergence. \textit{SIAM Journal on Control and Optimization, 61}(2), 789--823. 

\bibitem{WZZ2020}
Wang, H., Zariphopoulou, T., \& Zhou, X. Y. (2020). Reinforcement learning in continuous time and space: A stochastic control approach. \textit{Journal of Machine Learning Research, 21}, 198--1--198--34.


}
\end{thebibliography}
\end{document}